\documentstyle[12pt]{article}

\begin{document}

\topmargin 0pt
\oddsidemargin 7mm
\headheight 0pt
\topskip 0mm
\addtolength{\baselineskip}{0.40\baselineskip}

\hfill SOGANG-HEP 204/95

\hfill December 1995

\vspace{1cm}

\begin{center}
{\large \bf BFV-BRST Quantization of the Proca Model
            based on the Batalin-Tyutin formalism}
\end{center}

\vspace{1cm}

\begin{center}
Sean J. Yoon$^{1,2}$\footnote{Electronic address:
YOONSJ@GSCRL.goldstar.co.kr},
Yong-Wan Kim$^1$\footnote{Electronic address:
ywkim@physics.sogang.ac.kr},
and Young-Jai Park$^1$\footnote{Electronic address:
yjpark@physics.sogang.ac.kr}\\
\end{center}
\begin{center}
{\it \ $^1$ Department of Physics and Basic Science Research Institute \\
Sogang University, C.P.O. Box 1142, Seoul 100-611, Korea} \\
and\\
{\it \ $^2$ LG Electronics Research Center, Seoul 137-140, Korea}
\end{center}

\vspace{2cm}

\begin{center}
{\bf ABSTRACT}
\end{center}

We apply the Batalin-Tyutin Hamiltonian method to the abelian
Proca model in order to convert a second class constraint
system into a first class one systemetically by introducing the new field.
Then, according to the BFV formalism we obtain that the desired resulting
Lagrangian preserving standard
BRST sysmmetry naturally includes the well-known
St\"uckelberg scalar related to the explicit gauge-breaking effect due
to the presence of the mass term. Furthermore, we also discuss
the nonlocal symmetry structure of this model in the context of the
nonstandard BRST symmetry.

\newpage

\begin{center}
\section{\bf Introduction}
\end{center}

The Dirac method has been widely used in the Hamiltonian formalism [1]
to quantize second class constraint system, which does not form a closed
constraint algebra in Poisson brackets.
However, since the resulting Dirac brackets are generally field-dependent
and nonlocal, and have a serious ordering problem
between field operators, these are under unfavorable circumstances
in finding canonically conjugate pairs.
On the other hand,
the quantization of first class constraint system [2,3] has been well
appreciated in a gauge invariant manner preserving
Becci-Rouet-Stora-Tyutin (BRST) symmetry [4,5].  If second class constraint
system can be converted into first class one in an extended phase space,
we do not
need to define Dirac brackets and then the remaining quantization
program follows the method of Ref. [2-5].
This procedure has been extensively studied by Batalin, Fradkin, and Tyutin
[6,7] in the canonical formalism, and applied to various models [8-10]
obtaining the Wess-Zumino (WZ) action [11,12].

Recently, Banerjee [13] has applied the
Batalin-Tyutin (BT) Hamiltonian method [7]
to the second class constraint system of
the abelian Chern-Simons (CS) field theory [14-16], which yields first class
constraint algebra in an extended phase space
by introducing new fields.
As a result, he has obtained the new type of an abelian WZ action,
which cannot be obtained in the usual path-integral framework.
Very recently, we have quantized several interesting models [17] as well as
the nonabelian CS case [18], which yield the
weakly involutive first class system originating from the second class one,
by generalizing this BT formalism [7,13].
As shown in these works,
the nature of the second class constraint algebra originates from the
symplectic structure of CS term, not due to the local gauge symmetry
breaking.
Banerjee and Ghosh [19] have also considered a massive Maxwell theory,
which has the explicit gauge-breaking term, in the BT approach.
However, all these analyses do not carry out the covariant gauge fixing
procedure preserving the BRST symmetry.
On the other hand, Lavelle and McMullan (LM) recently have found
that QED exhibits a new nonlocal symmetry [20]. Several authors [21,22] have
extensively studied following their works.

In the present paper, we shall apply the BT Hamiltonian method [7]
to the Abelian Proca theory revealing the St\"uckelberg effect [23].
In section 2, we apply this formalism to the abelian Maxwell (Proca)
theory in order to systematically
convert a second class constraint system into a first
class one by introducing a new auxilary field $\rho$.
In section 3, we will briefly
discuss the special unitary gauge fixing reproducing the original second class
theory.
In section 4, we show that
by identifying this unphysical new field $\rho$ with the
St\"uckelberg scalar we naturally derive the St\"uckelberg scalar term related
to the explicit gauge-breaking mass term through
a standard BRST invariant gauge
fixing procedure according to the
Batalin-Fradkin-Vilkovisky (BFV) formalism.
We also analyse the nonlocal symmetry structure, which exists in QED,
of the Proca model in the context of the nonstandard BRST symmetry.

\vspace{1cm}
\begin{center}
\section {\bf The BT Formalism}
\end{center}

Now, we first apply the BT formalism to
the abelian massive Maxwell theory in four dimensions, whose dynamics are given
by
\begin{equation}
S = \int d^4x~
             [ -\frac{1}{4} F_{\mu\nu} F^{\mu\nu}
             + \frac{1}{2} m^2 A_\mu A^\mu ],
\end{equation}
where $F_{\mu\nu} = \partial_\mu A_\nu - \partial_\nu A_\mu$, and
$g_{\mu\nu} = diag(+,-,-,-)$.

The canonical momenta of gauge fields are given by
\begin{eqnarray}
\pi_0 &=& 0, \nonumber\\
\pi_i &=& {\dot A}^i + \partial_i A^0.
\end{eqnarray}
Then, $\Omega_1 \equiv \pi_0$ is a primary constraint [1].
The canonical Hamiltonian is
\begin{equation}
H_c = \int d^3x \left[
                \frac{1}{2} \pi_i^2 + \frac{1}{4} F_{ij} F^{ij}
              + \frac{1}{2} m^2 ((A^0)^2 + (A^i)^2)
              - A_0\Omega_2
                                                              \right],
\end{equation}
where $\Omega_2$ is the Gauss' law constraint,
which comes from the time evolution of $\Omega_1$, defined by
\begin{equation}
\Omega_2 = \partial^i \pi_i + m^2 A^0.
\end{equation}
The time evolution of the Gauss' law constraint generates no more
independent constraints. As a result, the full constraints of this model are
$\Omega_1$ and $\Omega_2$.
Then, they consist of the second class constraint algebra as follows
\begin{equation}
\Delta_{ij}(x,y) \equiv \{ \Omega_i(x), \Omega_j(y) \}
               = - m^2 \epsilon_{ij} \delta^3(x-y)
               ~~~~~(i, j = 1, 2),
\end{equation}
where we denote
$\epsilon_{12}=\epsilon^{12}=1$.

We now introduce new auxiliary fields $\Phi^i$ to convert the second
class constraint $\Omega_i$ into first class one in the extended phase space,
and assume that the Poisson algebra of the new fields is given by
\begin{equation}
   \{ \Phi^i(x), \Phi^j(y) \} = \omega^{ij}(x,y),
\end{equation}
where $\omega^{ij}$ is an antisymmetric matrix.
According to the BT method [7], the modified constraint in the
extended phase space is given by the polynomials of the auxiliary fields
$\Phi^i$ as follows
\begin{equation}
  \tilde{\Omega}_i(\pi_\mu, A^\mu, \Phi^i)
         =  \Omega_i + \sum_{n=1}^{\infty} \Omega_i^{(n)};
                       ~~~~~~\Omega_i^{(n)} \sim (\Phi^i)^n,
\end{equation}
satisfying the boundary condition,
$\tilde{\Omega}_i(\pi_\mu, A^\mu, \Phi^i = 0) = \Omega_i$.
The first order correction term in the infinite series [7] is given by
\begin{equation}
  \Omega_i^{(1)}(x) = \int d^3 y X_{ij}(x,y)\Phi^j(y),
\end{equation}
and the first class  constraint algebra of $\tilde{\Omega}_i$ requires the
condition as follows,
\begin{equation}
   \triangle_{ij}(x,y) +
   \int d^3 w~ d^3 z~
        X_{ik}(x,w) \omega^{kl}(w,z) X_{jl}(z,y)
         = 0.
\end{equation}
As was emphasized in Ref. [13,17], there is a natural arbitrariness
in choosing $\omega^{ij}$ and $X_{ij}$ from Eq. (6) and Eq. (8),
which corresponds to canonical transformation
in the extended phase space [6,7].
Thus, without any loss of generality, we take the simple solutions as
\begin{eqnarray}
  \omega^{ij}(x,y)
         &=&  \epsilon^{ij} \delta^3(x-y), \nonumber  \\
  X_{ij}(x,y)
         &=&  m \delta_{ij} \delta^3(x-y),
\end{eqnarray}
which are compatible with Eq. (9) as it should be.
Then, the modified constraint,
$\tilde{\Omega}_i$ give a strongly first class  constraint algebra,
\begin{equation}
  \{ \tilde{\Omega}_i(x), \tilde{\Omega}_j(y) \} = 0,
\end{equation}
where
\begin{equation}
\tilde{\Omega}_i=\Omega_i + m \Phi^i
\end{equation}
are the modified constraints including the auxiliay fields
$\Phi^i$ in the extended phase space.

Next, we derive the corresponding involutive Hamiltonian
in the extended phase space.
It is given by the infinite series [7],
\begin{equation}
 \tilde{H} = H_c + \sum_{n=1}^{\infty} H^{(n)};
{}~~~~~H^{(n)} \sim (\Phi^i)^n,
\end{equation}
satisfying the initial condition,
$\tilde{H}(\pi_\mu, A^\mu, \Phi^i = 0) = H_c$.
The general solution [7] for the involution of $\tilde{H}$ is given by
\begin{equation}
  H^{(n)} = -\frac{1}{n} \int d^3 x d^3 y d^3 z~
              \Phi^i(x) \omega_{ij}(x,y) X^{jk}(y,z) G_k^{(n-1)}(z)
              ~~~(n \geq 1),
\end{equation}
where the generating functions $G_k^{(n)}$ are given by
\begin{eqnarray}
  G_i^{(0)} &=& \{ \Omega_i^{(0)}, H_c \},  \nonumber  \\
  G_i^{(n)} &=& \{ \Omega_i^{(0)}, H^{(n)} \}_{\cal O}
                    + \{ \Omega_i^{(1)}, H^{(n-1)} \}_{\cal O}
                                       ~~~ (n \geq 1),
\end{eqnarray}
where the symbol ${\cal O}$ in Eq. (15) represents
that the Poisson brackets are calculated among the original variables, i.e.,
${\cal O}=(\pi_\mu, A^\mu)$.
Here, $\omega_{ij}$ and $X^{ij}$ are the inverse matrices of $\omega^{ij}$
and $X_{ij}$, respectively. Explicit calculations yield
\begin{eqnarray}
G_1^{(0)} &=& \Omega_2, \nonumber \\
G_2^{(0)} &=& m^2 \partial_i A^i,
\end{eqnarray}
which are substituted in (14) to obtain $H^{(1)}$,
\begin{equation}
 H^{(1)} =  \int d^3x \left[
                  m (\partial_i A^i) \Phi^1
                  - \frac{1}{m} (\partial^i \pi_i + m^2 A^0) \Phi^2
                  \right].
\end{equation}
This is inserted back in Eq. (15) in order to deduce $G_i^{(1)}$ as follows
\begin{eqnarray}
G_1^{(1)} &=& m \Phi^2, \nonumber\\
G_2^{(1)} &=& m \partial_i \partial^i \Phi^1.
\end{eqnarray}
Then, we obtain
$H^{(2)}$ by substituting $G_i^{(1)}$ in Eq. (14)
\begin{equation}
 H^{(2)}  =  \int  d^3x \left[
                  - \frac{1}{2} (\Phi^2)^2
                  - \frac{1}{2} (\partial_i \Phi^1)
                    (\partial^i \Phi^1)
                  \right].
\end{equation}
Finally, since
\begin{equation}
G_i^{(n)} = 0~~~~~~( n \geq 2),
\end{equation}
we obtain the complete form of the Hamiltonian ${\tilde{H}}$
after the $n=2$ finite truncations as follows
\begin{equation}
\tilde H = H_c + H^{(1)} + H^{(2)},
\end{equation}
which, by construction, is strongly involutive,
\begin{equation}
\{\tilde{\Omega}_i, \tilde H \} = 0.
\end{equation}
This  completes  the  operatorial conversion of the original
second class system with Hamiltonian $H_c$ and constraints $\Omega_i$
into the first class with Hamiltonian $\tilde H$ and constraints
$\tilde{\Omega}_i$.
{}From Eqs. (11) and (22),
one can easily see that the original second class constraint system
is converted into the first class system if one introduces two fields,
which are conjugated with each other in the extended phase space.
Note that the origin of second class constraint
is due to the explicit gauge symmetry breaking
term in the action (1).

Next we consider the partition function of the model
in order to present the Lagrangian corresponding to $\tilde{H}$
in the canonical Hamiltonian formalism.
As a result, we will unravel the correspondence of
the Hamiltonian approach with
the well-known St\"uckelberg's formalism.
First, let us identify the new variables
$\Phi^i$ as a canonically conjugate pair
($\rho$, $\pi_\rho$) in the Hamiltonian formalism, {\it i.e.,}
\begin{equation}
\Phi^i = ( m \rho,~
          \frac{1}{m}\pi_\rho )
\end{equation}
satisfying Eqs. (6) and (10).
Then, the starting phase space partition function is given by the Faddeev
formula [3,24] as follows
\begin{equation}
Z=  \int  {\cal D} A^\mu
          {\cal D} \pi_\mu
          {\cal D} \rho
          {\cal D} \pi_\rho
               \prod_{i,j = 1}^{2} \delta(\tilde{\Omega}_i)
                           \delta(\Gamma_j)
                det \mid \{ \tilde{\Omega}_i, \Gamma_j \} \mid
                e^{iS},
\end{equation}
where
\begin{equation}
S  =  \int d^4x \left(
           \pi_\mu {\dot A}^\mu + \pi_\rho {\dot \rho} - \tilde {\cal H}
            \right),
\end{equation}
with Hamiltonian density $\tilde {\cal H}$ corresponding to Hamiltonian
$\tilde H$ (21), which is now expressed in terms
of $(\rho, \pi_\rho)$ instead of $\Phi^i$.
The gauge fixing conditions $\Gamma_i$ are chosen
so that the determinant occurring in
the functional measure is nonvanishing.
Moreover, $\Gamma_i$ may be assumed to be independent of the momenta
so that these are considered as Faddeev-Popov type gauge conditions [24].

Before performing the momentum integrations to obtain the
partition function in the configuration space,
it seems appropriate to comment on the involutive Hamiltonian.
If we directly use the Hamiltonian (21) following the previous analysis done
by Banerjee {\it {et al.}} [19],
we will finally obtain the non-local action corresponding to this Hamiltonian
due to the existence of $(\partial^i \pi_i)^2$--term in the action
when we carry out the functional integration over $\pi_{\rho}$ later.
Furthermore, if we use this Hamiltonian,
we can not also naturally generate
the first class Gauss' law constraint $\tilde{\Omega}_2$ from
the time evolution of the primary constraint $\tilde{\Omega}_1$,
which is the first class.
Therefore, in order to avoid these serious problems,
we use the equivalent first class Hamiltonian
without any loss of generality,
which only differs from the involutive Hamiltonian (21)
by adding a term proportional to the first class constraint
$\tilde{\Omega}_2$ as follows
\begin{equation}
\tilde{H}' = \tilde H + \frac{\pi_\rho}{m^2} \tilde{\Omega}_2.
\end{equation}
Then, we have the natural first constraint system such that
\begin{equation}
\{ \tilde{\Omega}_1, \tilde{H}'\} = \tilde{\Omega}_2, ~~~
\{ \tilde{\Omega}_2, \tilde{H}'\} = 0.
\end{equation}
Note that when we act this modified Hamiltonian (26) on physical states,
the difference is trivial because such states are
annihilated by the first class constraints.
Similarly, the equations of motion for observable ({\it i.e.}
gauge invariant variables) will also be unaffected by this difference
since $\tilde{\Omega}_2$ can be regarded as
the generator of the gauge transformations.

\vspace{1cm}
\begin{center}
\section {\bf The Original Unitary Gauge Fixing}
\end{center}

Now, we consider the following effective phase space partition function
\begin{eqnarray}
Z &=& \int {\cal D} \pi_{\mu} {\cal D} A^{\mu}
           {\cal D} \pi_\rho
           {\cal D} \rho
           \prod^2_{i,j = 1}
            \delta(\tilde{\Omega}_i) \delta(\Gamma_j)
            det \mid \{ \tilde{\Omega}_i,  \Gamma_j \} \mid
            e^{iS'}, \nonumber \\
S' &=& \int  d^4 x ~( \pi_{\mu} {\dot A^{\mu}} +
            \pi_{\rho} \dot{\rho}
            -  \tilde{\cal H'}).
\end{eqnarray}
The trivial $\pi_0$ integral is performed by exploiting the delta function
$\delta(\tilde{\Omega}_1) = \delta(\pi_0 + m^2 \rho)$ in (28).
On the other hand,
the other delta function $\delta(\tilde{\Omega}_2) = \delta
(\partial^i \pi_i + m^2 A^0 + \pi_\rho)$ can be expressed
by its Fourier transform with Fourier variable $\xi$ as follows
\begin{equation}
\delta(\tilde{\Omega}_2) = \int {\cal D}\xi
                          e^{-i \int d^4x ~\xi \tilde{\Omega}_2}.
\end{equation}
Making a change of variable $A^0 \to A^0 + \xi$, we obtain the action
\begin{eqnarray}
S &=& \int d^4x~ [ \pi_i {\dot A}^i - m^2 \rho ( \dot{A^0} + \dot{\xi} )
                    + \pi_{\rho}\dot{\rho}
                - \frac{1}{2} \pi_i^2 - \frac{1}{4} F_{ij} F^{ij}
                + \frac{1}{2} m^2 (A^0)^2
                + \frac{1}{2} m^2 A_i A^i \nonumber\\
   &+& A^0 \partial^i \pi_i
                 - m^2 \partial_i A^i \rho
                - \frac{1}{2m^2} \pi_\rho^2
                + \frac{1}{2} m^2 \partial_i \rho \partial^i \rho
                - \xi \pi_\rho - \frac{1}{2} m^2 \xi^2 ],
\end{eqnarray}
where the corresponding measure is given by
\begin{equation}
[{\cal D} \mu] = {\cal D} \xi
                 {\cal D} \pi_i
                 {\cal D} A^\mu
                {\cal D} \pi_\rho
                {\cal D} \rho
                \prod_j
         \{ \delta[ \Gamma_j (A^0+\xi, A^i, \pi_i, \rho, \pi_\rho) ] \}
               det \mid \{\tilde{\Omega}_i, \Gamma_j \} \mid.
\end{equation}
Performing the Gaussian integral over $\pi_i$, this yields the action
as follows
\begin{eqnarray}
S_u &=&  \int d^4x~ \left[ - \frac{1}{4} F_{\mu\nu} F^{\mu\nu}
            + \frac{1}{2} m^2 A_\mu A^\mu
 + \pi_\rho (\dot{\rho}-\xi-\frac{1}{2m^2}\pi_\rho) \right.
\nonumber \\
    &-& \left. m^2 \rho ( \dot{A^0} + \dot{\xi} )
           - m^2 \partial_i A^i \rho
           + \frac{1}{2} m^2 \partial_i\rho \partial^i\rho
           - \frac{1}{2} m^2 \xi^2 \right].
\end{eqnarray}
Now, we choose the unitary gauge as follows
\begin{equation}
\Gamma_i = ( \rho,~ \pi_\rho ).
\end{equation}
Note that this gauge fixing is consistent because when we take the
gauge fixing condition $\rho = 0$, another condition $\pi_\rho = 0$ is
naturally generated from the time evolution of $\rho$, $i.e.$,
$\dot{\rho} = \{\rho , H_u \}=-\frac{1}{m^2}\pi_\rho = 0$,
where the Hamiltonian $H_u$ corresponds to the intermediate action $S_u$.
However, we can not smoothly choose the massless limit in this unitary
gauge because $\dot{\rho}$ tends to infinity for this limit.
In this gauge, we get
\begin{equation}
\{ \tilde{\Omega}_i(x), \Gamma_j (y) \}
=\epsilon_{ij}\delta^3(x-y).
\end{equation}
Then, we easily recover the original system. Therefore, we can interpret
the original system (1) as a gauge-fixed version of the extended
gauge system (13), (21), and (26).

\vspace{1cm}
\begin{center}
\section {\bf The BFV-BRST Gauge Fixing}
\end{center}

In this section, we first briefly recapitulate
the BFV formalism [2,3] which is applicable for the general theories
with first-class constraints.
For simplicity, this formalism is restricted to a finite
number of phase space variables. This makes the discussion simpler and
conclusions more apparent.

First of all, consider a phase space of canonical variables $q^{i},~p_{i}$
($i$ = 1, 2, $\cdots$, n) in terms of which the canonical Hamiltonian
$H_{c}(q^{i},p_{i})$ and the constraints $\Omega_{a}(q^{i},p_{i}) \approx 0$
($a$ = 1, 2, $\cdots$, m) are given. We assume that the constraints
satisfy the constraint algebra [2,3]
\begin{eqnarray}
\left[~\Omega_{a}, \Omega_{b} ~\right] &=& i\Omega_{c} U^{c}_{a b}~,
           \nonumber \\
\left[~H_{c}, \Omega_{a} ~\right] &=& i\Omega_{b} V^{b}_{a}~,
\end{eqnarray}
where the structure coefficients $U^{c}_{a b}$ and $V^{b}_{a}$ are
functions of the canonical variables. We also assume that the constraints
are irreducible, which means that locally
there exists an invertible change of
variables such that $\Omega_{a}$ can be identified with the $m$-unphysical
momenta.

In order to single out the physical variables, we can introduce the
additional conditions $\Phi^{a}(q^{i},p_{i}) \approx 0$ with
$\det [ \Phi^{a}, \Omega_{b} ] \neq 0$ at least in the vicinity
of the constraint surface $\Phi^{a} \approx 0$ and $\Omega_{a} \approx 0$.
Then, $\Phi^a$ play the roles of gauge-fixing functions. That is to say,
from the condition of the time stability of the constraints, there exists
a family of phase space trajectories. By selecting one of these
trajectories through the conditions of $\Phi^{a} \approx 0$, we can get
the 2($n-m$) dimensional physical phase space
noted by $q^{*}$ and $p^{*}$ [1-3].
And then, $\Phi^{a} (q^{i},p_{i})$ can be
identified with the $m$-unphysical coordinates.

The described dynamical system
\begin{equation}
{\cal Z} = \int \!  [ d q^{i} d p_{i} ]~ \delta ( \Omega_{a} )
             \delta ( \Phi^{b} ) det \mid [ \Phi^{b}, \Omega_{a} ] \mid~
             {\mbox {\large e}}^{ i\!\!\int \!\! d x(p \dot{q} - H_{c})}
\end{equation}
is completely equivalent to an effective quantum theory
only depending on the physical canonical variables $q, p$ of the physical
phase space [2,3].
Note that the constraints $\Omega_{a} \approx 0$ and $\Phi^{a} \approx 0$
together with the Hamilton equations may be obtained from a action
\begin{equation}
S~=~\int \! d t ~(p_{i} \dot{q}^{i}-H_{c}-N^{a} \Omega_{a}-B_{a}
   \Phi^{a} )~,
\end{equation}
where $N^{a}$ and $B_{a}$ are Lagrange
multiplier fields canonically conjugated to each other, obeying the
commutation relations
\begin{equation}
[~N^{a},
                B_{b}~]~=~i\delta^{a}_{b}~,
\end{equation}
and the gauge-fixing
conditions contain $\lambda^{a}$ in the following general form
\begin{equation}
   \Phi^{a}~=~\dot{N}^{a}~+~\chi^{a}(q^{i},p_{i},N^{a})~,
\end{equation}
where $\chi^{a}$ are arbitrary functions.
Furthermore, we can see
that the Lagrange multiplier $N^{a}$ become dynamically active, and
$B_{a}$ serve as their conjugate momenta. This consideration naturally
leads to the canonical formalism in an extended phase space.

In order to make the equivalence to the initial theory with constraints
in the reduced phase space,
we may introduce two sets of canonically conjugate, anticommuting ghost
coordinates and momenta ${\cal C}^{a},~\overline{\cal P}_{a}$ and
${\cal P}^{a},~\overline{\cal C}_{a}$ such that
\begin{equation}
[ {\cal C}^{a}, \overline{\cal P}_{b} ]~=~[ {\cal P}^{a},
\overline{\cal C}_{b} ]~=~i\delta^{a}_{b}~.
\end{equation}
The quantum theory is defined by the extended phase space functional
integral
\begin{equation}
{\cal Z}_{\Psi}~=~ \int
         [d\mu]
         {\mbox{\large e}}^{iS_{\Psi}}~,
\end{equation}
where the action is now
\begin{equation}
S_{\Psi}~=~\int \! d t ~\{p_{i} \dot{q}^{i}~+~B_{a} \dot{N}^{a}~+~
           {\overline{\cal P}}_{a} \dot{\cal C}^{a}~+~
           {\overline{\cal C}}_{a} \dot{\cal P}^{a} ~-~
           H_{m}~+~i[Q,\Psi] \},
\end{equation}
and $[d\mu]$ is the Liouville measure, {i.e.,}
\begin{equation}
[d\mu]=[dq^i dp_i dN^a dB_a d{\cal C}^a d\overline{\cal P}_a
d{\cal P}^a d\overline{\cal C}_a],
\end{equation}
on the constraint phase space.
Here, the BRST-charge $Q$ and the fermionic gauge-fixing function $\Psi$
are defined by
\begin{eqnarray}
Q &=& {\cal C}^{a}\Omega_{a}~-~\frac{1}{2} {\cal C}^{b} {\cal C}^{c}
    U^{a}_{c b} {\overline{\cal P}}_{a}~+~{\cal P}^{a} B_{a}~, \nonumber \\
\Psi &=& {\overline{\cal C}}_{a} \chi^{a}~+~ {\overline{\cal P}}_{a} N^{a}~,
\end{eqnarray}
respectively.
$H_{m}$ is the BRST invariant Hamiltonian, called the minimal Hamiltonian,
\begin{equation}
H_{m}~=~H_{c} + {\overline{\cal P}}_{a} V^a_b {\cal C}_b.
\end{equation}

Now, in order to derive a BRST gauge-fixed covariant action
for the abelian Proca model considered in the previous section,
let us introduce the ghosts and anti-ghosts
along with auxiliary fields
$({\cal C}^i,~ {\overline{\cal P}}_i),~({\cal P}^i,
{}~{\overline{\cal C}}_i ),
{}~(N^i , B_i)$,
where $i = 1,~2$, according to the above BFV formalism in the
extended phase space.
The nilpotent BRST-charge $Q$,
the fermionic gauge-fixing function $\Psi$, and the minimal
Hamiltonian $H_m$ have the following concrete forms
\begin{eqnarray}
Q~&=&~\int\!d^3x~[~{\cal C}^i \tilde{\Omega}_i ~+~
                  {\cal P}^i B_i ~], \nonumber \\
\Psi~&=&~\int\!d^3x~[~{\overline{\cal C}}_i \chi^i
                    ~+~
                     {\overline{\cal P}}_i N^i ~], \nonumber \\
H_{m}~&=&~\tilde{H}'~-~\int\!d^3x
          ~[\overline{\cal P}_{2} {\cal C}^{1}],
\end{eqnarray}
where
\begin{equation}
\chi^1 =A^0 ,~~~\chi^2 = \partial_i A^i + \frac{\textstyle \alpha}
{\textstyle 2}B_2,
\end{equation}
as gauge fixing functions, and $\alpha$ is an arbitrary parameter.
Note that the form of $H_{m}$ is simpler than that in Ref.[8] due
to our improved Hamiltonian (26) proposed in our previous works [17].

The BRST-charge $Q$,
the fermionic gauge-fixing function $\Psi$, and the minimal
Hamiltonian $H_m$ satisfy
the following relations,
\begin{eqnarray}
i \left[Q, H_{m} \right]&=& 0, \nonumber \\
Q^{2}~=~\left[Q,Q\right]&=& 0, \nonumber \\
\left[~\left[ \Psi,Q\right], Q\right]&=&0~,
\end{eqnarray}
where they are the conditions of physical subspace being imposed.

Then, the effective action is
\begin{equation}
S_{eff} = \int\!d^4\!x~[~\pi_0 \dot{A}^0 + \pi_i \dot{A}^i
                    + \pi_\rho \dot{\rho}
                    + B_2\dot{N}^2+ \overline{\cal P}_i\dot{\cal C}^i
                    + \overline{\cal C}_2 \dot{\cal P}^2~ ] - H_{total},
\end{equation}
where \( H_{total} = H_{m} - i[Q,\Psi] \).
Note that we could suppress the term $\int d^4x ( B_{1} \dot{N}^{1} +
\overline{\cal C}_{1} \dot{\cal P}^1) = -i [ Q, \int d^4x \overline{C}_1
\dot{N}^1 ]$ in the Legendre transformation by replacing $\chi^1$ with
$\chi^1+\dot{N}^1$.

\vspace{1cm}
\begin{center}
\subsection {\bf The Standard Local Gauge Fixing}
\end{center}

The fields $B_1 ,~N^1 ,~\overline{\cal C}_1 , {\cal P}^1 ,
{}~\overline{\cal P}_1 , {\cal C}^1$ , and $A^0$ are eliminated,
and integration of $\pi_0$ gives the delta functional by  using of
Gaussian integration. Then we obtain the generating functional as follows
\begin{equation}
 {\cal Z} = \int [{\cal D}\mu] \exp [i S_{eff}],
\end{equation}
where the effective action is
\begin{eqnarray}
S_{eff}&=&\int\!d^4x~
      [~\pi_i \dot{A}^i
      + \pi_{\rho} \dot{\rho}
      +B\dot{N}
      +{\overline{\cal P}}\dot{\cal C}
      +{\overline{\cal C}}\dot{\cal P}
   -\frac{1}{2}(\pi^i)^2
       - \frac{1}{2m^2}(\pi_\rho)^2  \nonumber \\
    &&~~ - \frac{1}{4} F_{ij} F^{ij}
       -\frac{1}{2} m^2 (A^i)^2
       - m^2 \rho \partial_i A^i
       + \frac{1}{2} m^2 \partial_i \rho \partial^i \rho  \nonumber \\
    &&~~ +N(\partial_i\pi^i + \pi_\rho)
       +B(\partial_iA^i + \frac{1}{2}\alpha B)
       +\partial_i{\overline{\cal C}}
             \partial^i{\cal C} + {\overline{\cal P}}{\cal P}~],
\end{eqnarray}
and the Liouville measure of the extended phase space is given by
\begin{equation}
[{\cal D}\mu] = [{\cal D}\pi_i ~{\cal D}A^i
                 ~{\cal D}\rho ~{\cal D}\pi_{\rho}
                 ~{\cal D}B ~{\cal D}N
                 ~{\cal D}\overline{\cal P} ~{\cal D}C
                 ~{\cal D}\overline{\cal C} ~{\cal D}{\cal P} ].
\end{equation}
Here we have redefined
$N^2 \equiv N,~B_2 \equiv B,~ {\overline{\cal C}}_2 \equiv{\overline{\cal C}},~
{\cal C}^2 \equiv{\cal C},~ {\overline{\cal P}}_2\equiv{\overline{\cal P}}$,
and
 ${\cal P}^2 \equiv{\cal P}.$
Performing the integrations of $\pi^{i},~\pi_{\rho},~{\cal P}$, and
${\overline{\cal P}}$,
and identifying with $N=-A^0$, we get the
following covariant effective action
\begin{eqnarray}
S_{eff}=\int\!d^4x~[ -\frac{1}{4}F_{\mu \nu}F^{\mu \nu}
            + \frac{1}{2} m^{2}
               ( A_{\mu} + \partial_\mu \rho )^2
 - A^{\mu}\partial_{\mu}B + \frac{1}{2}\alpha (B)^2
             - \partial_{\mu}{\overline{\cal C}}\partial^{\mu}{\cal C}~],
\end{eqnarray}
which is invariant under the BRST transformation
\begin{eqnarray}
\delta_{B}A_{\mu} &=& -\lambda\partial_{\mu}{\cal C},
            ~~~~~~\delta_{B} \rho = \lambda {\cal C},  \nonumber \\
 \delta_{B}{\cal C}  &=& 0,
          ~~~~~~\delta_{B}{\overline{\cal C}} = - \lambda B,
          ~~~~~~\delta_{B}B = 0,
\end{eqnarray}
where $\lambda$ is a constant Grassmann parameter, and
the corresponding final measure is given by
\begin{equation}
[{\cal D}\mu]=[{\cal D} A^\mu ~{\cal D} \rho ~{\cal D} B
               ~{\cal D} C ~{\cal D} \overline{\cal C}].
\end{equation}
Therefore, in Eq. (53) we see that the auxiliary BF field $\rho$ is exactly
the well-known St\"uckelberg scalar [23].
Note that in this gauge, we can smoothly choose the massless limit
because the gauge fixing conditions (47)
are independent of mass $m^2$, and then
obtain the well-known QED result.
On the other hand,
this BRST symmetry gives a conserved current as
\begin{equation}
J_{B\mu}=F_{\mu\nu}\partial^\nu {\cal C} + m^2 (A_\mu + \partial_\mu \rho)
         {\cal C} + B\partial_\mu{\cal C},
\end{equation}
through Noether's theorem.

\vspace{1cm}
\begin{center}
\subsection {\bf The Nonstandard Nonlocal Gauge Fixing}
\end{center}

Consider the BFV formalism in the previous section up to the point, where
the integration over the momentum $\pi_{\rho}$ was performed and
the following action was obtained.
\begin{eqnarray}
S_{eff}&=&\int\!d^4x~\left[ -\frac{1}{4}F_{\mu \nu}F^{\mu \nu}
            + \frac{1}{2} m^{2}
               ( A_{\mu} + \partial_\mu \rho )^2
       - A^{\mu}\partial_{\mu}B + \frac{1}{2}\alpha B^2 \right. \nonumber \\
   &&  \left. - \partial_{i}{\overline{\cal C}}\partial^{i}{\cal C}
 +{\overline{\cal P}}\dot{\cal C}
      +{\overline{\cal C}}\dot{\cal P}
      + {\overline{\cal P}}{\cal P}~\right].
\end{eqnarray}
This action is invariant under the BRST transformation,
which have the form
\begin{eqnarray}
\delta_{B}A_{0} &=& -\lambda{\cal P},
{}~~~~~~\delta_{B}A_{i} = -\lambda\partial_{i}{\cal C},
{}~~~~~~\delta_{B} \rho = \lambda {\cal C},  \nonumber \\
 \delta_{B}{\cal C}  &=& 0,
          ~~~~~~\delta_{B}{\overline{\cal C}} = - \lambda B,
          ~~~~~~\delta_{B}B = 0, \nonumber \\
 \delta_{B}{\cal P}  &=& 0,
          ~~~~~~\delta_{B}{\overline{\cal P}} = - \lambda [
          -\partial_{i}F^{0i} + m^{2} (\dot{\rho} + A^{0})] .
\end{eqnarray}

Now, if we perform the integration
over the ghost fields instead of their momenta,
we can also find the nonlocal symmetry structure
in the Proca theory as well as QED [20]. First, performing the
integration over the ghost field ${\cal C}$, we get the following
delta function
\begin{equation}
\delta( \partial_{i}\partial^{i}{\overline{\cal C}}
        -\dot{{\overline{\cal P}}} )~=~det(\partial_{i}\partial^{i})
  \delta( {\overline{\cal C}}-\frac{1}{\partial_{i}\partial^{i}}
  \dot{{\overline{\cal P}}}).
\end{equation}
Next, performing the integration over ${\overline{\cal C}}$,
we get non-local ghost Lagrangian
\begin{equation}
S_{gh} ~=~\int\!d^4x~[ \dot{{\overline{\cal P}}}
   \frac{1}{\partial_{i}\partial^{i}}
   \dot{{\cal P}} -{\overline{\cal P}}{\cal P} ].
\end{equation}
Notice that the appearance of the nonlocal term in the ghost action has
a result of this unusual integration.
However,
this form can be also simply obtained by the change of variables
\begin{equation}
{\cal C} ~\to~ \frac{1}{\partial_{i}\partial^{i}}{\overline{\cal P}} ,~~
{\overline{\cal C}} ~\to~{\cal P},
\end{equation}
in the Eq. (53). Under these replacements, we have the nonlocal BRST charge,
that we call $Q'$, being
\begin{equation}
Q'~=~\int\!d^3x~ \left[
     B{\overline{\cal C}}+{
     -\partial_{i}F^{0i} + m^{2} (\dot{\rho} + A^{0})}
 \frac{1}{\partial_{i}\partial^{i}}{\overline{\cal P}} \right].
\end{equation}
Then, the effective action is BRST invariant under the following
transformations as
\begin{eqnarray}
\delta_{B}'A_{\mu} &=& -\lambda\partial_{\mu}
                    (\frac{1}{\partial_{i}\partial^{i}}{\overline{\cal P}}) ,
 ~~~~~~\delta_{B}' \rho = \lambda \frac{1}{\partial_{i}\partial^{i}}
                    {\overline{\cal P}} ,  \nonumber \\
 \delta_{B}'{\overline{\cal P}}  &=& 0,
          ~~~~~~\delta_{B}'{\cal P} = - \lambda B,
          ~~~~~~\delta_{B}'B = 0.
\end{eqnarray}
This nonlocal BRST symmetry yields a conserved current through Noether's
theorem as follows
\begin{equation}
J_{B \mu}'~=~F_{\mu \nu}\partial^{\nu}
           \frac{1}{\partial_{i}\partial^{i}}{\overline{\cal P}}+
           m^{2} ( A_{\mu} + \partial_\mu \rho )
           \frac{1}{\partial_{i}\partial^{i}}{\overline{\cal P}} +
           B \partial_{\mu}
           \frac{1}{\partial_{i}\partial^{i}}{\overline{\cal P}}.
\end{equation}
Note that performing the change of variable (61),
these nonlocal symmetry and conserved current turn into just the original
local theory (56).

In conclusion, we have applied our improved Batalin-Tyutin method [17],
which systematically converts the second class system
into the first class one
by introducing the new auxiliary fields,
to the Proca theory.
According to the BFV formalism with the efficient first class Hamiltonian
through BT analysis, we have shown that the resulting
Proca Lagrangian preserving
standard BRST symmetry naturally includes the well-known St\"ukelberg
scalar needed for the anomaly free theory
by identifying this scalar with one of auxiliary fields.
Furthermore, we have
shown that the nonlocal symmetry structure recently proposed in QED
also exists in the Proca model through the nonstandard BRST gauge-fixing
procedure.

\vspace{1cm}

\begin{center}
\section*{Acknowledgements}
\end{center}

We would like to thank W. T. Kim and B. H. Lee for helpful discussions.
The present study was supported by
the Basic Science Research Institute Program,
Ministry of Education, Project No. 95-2414.

\newpage

\section*{References}

\begin{description}{}

\item{1.} P. A. M. Dirac: Lectures on quantum mechanics, New York:
            Yeshiba University Press 1964
\item{2.} E. S. Fradkin, G. A. Vilkovisky:
            Phys. Lett. 55B (1975) 224;
           I. A. Batalin, G. A. Vilkovisky:
            {\it ibid.,} 69B (1977) 309;
           E. S. Fradkin, T. E. Fradkina: {\it ibid.,}
           72B (1977) 343
\item{3.} M. Henneaux: Phys. Rep. C126 (1985) 1
\item{4.} C. Becci, A. Rouet, R. Stora:
           Ann. Phys. [N.Y.]  98 (1976) 287;
           I. V. Tyutin: Lebedev Preprint 39 (1975)
\item{5.} T. Kugo, I. Ojima:
           Prog. Theor. Phys. Suppl. 66 (1979) 1
\item{6.} I. A. Batalin, E. S. Fradkin:
           Nucl. Phys. B279 (1987) 514;
           Phys. Lett. B180 (1986) 157
\item{7.} I. A. Batalin, I. V. Tyutin:
           Int. J. Mod. Phys. A6 (1991) 3255
\item{8.} T. Fujiwara, Y. Igarashi, J. Kubo:
            Nucl. Phys. B341 (1990) 695;
	    O. Dayi: Phys. Lett. B210 (1988) 147
\item{9.} Y.-W. Kim,  S.-K. Kim, W. T. Kim, Y.-J. Park, K. Y. Kim,
          and Y. Kim:
            Phys. Rev. D46 (1992) 4574
\item{10.} R. Banerjee, H. J. Rothe, K. D. Rothe:
           Phys. Rev. D49 (1994) 5436
\item{11.} L. D. Faddeev, S. L. Shatashivili:
           Phys. Lett. B167 (1986) 225;
           O. Babelon, F. A. Shaposnik, C. M. Vialett:
           Phys. Lett. B177 (1986) 385;
           K. Harada, I. Tsutsui: Phys. Lett. B183 (1987) 311;
           J.-G. Zhou, Y.-G. Miao, Y.-Y. Liu: Mod. Phys. Lett.
           A9 (1994) 1273
\item{12.} J. Wess, B. Zumino: Phys. Lett. 37B (1971) 95
\item{13.} R. Banerjee: Phys. Rev. D48 (1993) R5467
\item{14.} Edited by S. Treiman et al.:
           Topological Investigations of
           Quantized Gauge Theories, Singapore:
           World Scientific 1985
\item{15.} G. Semenoff: Phys. Rev. Lett. 61 (1988) 517;
           G. Semenoff, P. Sodano: Nucl. Phys. B328 (1989) 753
\item{16.} R. Banerjee: Phys. Rev. Lett. 69 (1992) 17;
           Phys. Rev. D48 (1993) 2905
\item{17.} Y.-W. Kim, Y.-J. Park, K. Y. Kim, and Y. Kim:
           Phys. Rev. D51 (1995) 2943;
           J.-H. Cha, Y.-W. Kim, Y.-J. Park, Y. Kim,
           S.-K. Kim, and W. T. Kim:
           preprint hep-th/9507052
           (to appear Z. Phys. C, 1995)
\item{18.} W. T. Kim, Y. -J. Park: Phys. Lett. B336 (1994) 376
\item{19.} N. Banerjee, R. Banerjee, S. Ghosh: Ann. Phys.
           241 (1995) 237;
           N. Banerjee, S. Ghosh, R. Banerjee:
           Nucl. Phys. B417 (1994) 257;
           Phys. Rev. D49 (1994) 1996
\item{20.} M. Lavelle, D. McMullan: Phys. Rev. Lett. 71
           (1993) 3758
\item{21.} Z. Tang, D. Finkelstein: Phys. Rev. Lett. 73
           (1994) 3055;
           H. Yang, B.-H. Lee: J. Korean Phys. Soc. 28
           (1995) 572;
           D. K. Park, H. S. Kim, J. K. Kim: preprint hep-th/9511003;
           S. J. Rabello, P. Gaete: preprint hep-th/9504132
\item{22.} H. Shin, Y.-J. Park, Y. Kim, and W. T. Kim:
           preprint hep-th/9506166
\item{23.} E. C. G. St\"uckelberg: Helv. Phys. Act. 30 (1957) 209;
           L. D. Faddeev: Theor. Math. Phys. 1 (1970) 1
\item{24.} L. D. Faddeev, V. N. Popov: Phys. Lett. 25B (1967) 29
\end{description}

\end{document}